\documentclass[english,aps,preprint]{iopart}
\usepackage[T1]{fontenc}
\usepackage[latin9]{inputenc}
\usepackage{babel}

\usepackage{textcomp}
\usepackage{amstext}
\usepackage{graphicx}
\usepackage{amssymb}
\usepackage[unicode=true, pdfusetitle,
 bookmarks=true,bookmarksnumbered=false,bookmarksopen=false,
 breaklinks=false,pdfborder={0 0 1},backref=false,colorlinks=false]
 {hyperref}

\makeatletter

\newcommand{\lyxmathsym}[1]{\ifmmode\begingroup\def\b@ld{bold}
  \text{\ifx\math@version\b@ld\bfseries\fi#1}\endgroup\else#1\fi}

\usepackage{iopams}
\usepackage{setstack}

\makeatother

\begin{document}
This line only printed with reprint option

\title{Influence of B - site Disorder in $La_{0.5}Ca_{0.5}Mn_{1-x}B_{x}O_{3}$
(B = Fe, Ru, Al and Ga) Manganites}

\author{Indu Dhiman}

\address{Solid State Physics Division, Bhabha Atomic Research Centre, Mumbai
- 400\hspace*{0.05cm}085, India}

\author{A. Das}

\address{Solid State Physics Division, Bhabha Atomic Research Centre, Mumbai
- 400\hspace*{0.05cm}085, India}

\ead{adas@barc.gov.in}

\author{A. K. Nigam}

\address{Department of Condensed Matter Physics and Material Science, Tata Institute of Fundamental Research, Colaba, Mumbai - 400\hspace*{0.05cm}005,
India}

\author{Urs Gasser}

\address{Laboratory for Neutron Scattering, Paul Scherrer Institute, ETH Zurich,
CH-5232 Villigen PSI, Switzerland}

\begin{abstract}
We have investigated the influence of B - site doping on the crystal
and magnetic structure in $La_{0.5}Ca_{0.5}Mn_{1-x}B_{x}O_{3}$ (B
= Fe, Ru, Al and Ga) compounds using neutron diffraction, small angle
neutron scattering, magnetization and resistivity techniques. The
B - site doped samples are isostructural and possess an orthorhombic
structure in \textit{Pnma} space group at 300K. A structural transition
from orthorhombic to monoclinic is found to precede the magnetic transition
to CE - type antiferromagnetic state in few of these samples. On doping
with Fe, charge and orbitally ordered CE - type antiferromagnetic
state is suppressed, followed by the growth in ferromagnetic insulating
phase in $0.02\leq x\leq0.06$ compounds. At higher Fe doping in $x>0.06$,
the ferromagnetic state is also suppressed and no evidence of long
range magnetic ordering is observed. In Ru doped samples $(0.01\leq x\leq0.05)$,
the ferromagnetic metallic state is favored at $T{}_{C}\approx200K$
and $T_{MI}\approx125K$ and no significant change in $T_{C}$ and
$T_{MI}$ as a function of Ru doping is found. In contrast, with non
magnetic Al substitution for $0.01\leq x\leq0.03$, the charge ordered
CE - type antiferromagnetic state coexists with the ferromagnetic
metallic phase. With further increase in Al doping $(0.05\leq x\leq0.07)$,
both CE - type antiferromagnetic and ferromagnetic phases are gradually
suppressed. This behavior is accompanied by the evolution of A - type
antiferromagnetic insulating state. Eventually, at higher Al doping
$(0.10\leq x\leq0.13)$, this phase is also suppressed and signature
of spin glass like transition are evident in M(T). Likewise, substitution
with Ga is observed to induce similar effects as described for Al
doped samples. The presence of short ranged ferromagnetic ordering
has been further explored using small angle neutron scattering measurements
in few of the selected samples.
\end{abstract}



\maketitle

\section{Introduction}

Perovskite manganites $R{}_{1-x}A_{x}MnO_{3}$ (R: trivalent rare
earth ion and A: divalent alkaline earth ions, Ca, Ba and Sr) have
been widely studied due to the presence of variety of phenomena, viz.
colossal magnetoresistance, charge, orbital, and spin ordering and
phase separation behavior. In particular, the nature of charge, orbital
and spin ordering is strongly influenced by various perturbations
such as, disorder effects at the rare earth (A - site) and transition
metal (B - site) site, hydrostatic pressure, magnetic and electric
field, and by varying the particle size \cite{C. N. R. Rao,E. Dagotto,J. B. Goodenough,Y. Tokura,G. van Tendeloo}.

Several experimental and theoretical studies have been devoted
toward the investigations of substitutional disorder effects in
charge ordered manganites. Generally, two sources of
substitutional disorder have been described in charge ordered
manganites: A - site and B - site disorder. The magnetic and
transport properties in several systems are observed to be
systematically influenced by structural distortions and lattice
disorder associated with the average A - site ionic radii
$(<r{}_{A}>)$ and variance $(\sigma{}^{2})$ due to size mismatch
between the A - site cations \cite{C. N. R. Rao-1,A. Arulraj,I.
Dhiman,I. Dhiman-1,D. Akahoshi}. The B - site substitution, as
against A - site, induces disorder directly in the Mn - O network,
leading to much stronger impact on the magnetic and transport
properties. The high sensitivity of charge and orbitally ordered
antiferromagnetic state to B - site disorder has been studied in
several half doped $R{}_{0.5}A_{0.5}MnO_{3}$ manganites \cite{G.
van Tendeloo,Y. Tokura-1,M. Salamon,E. Dagotto-1}. These studies
show that a few percent of Mn - site doping brings about drastic
changes in magnetic and transport properties, without
significantly influencing the crystal structure. The B - site
dopants such as Cr, Ru, Ni, or Co in low bandwidth
$R{}_{0.5}Ca_{0.5}MnO_{3}$ (R = Nd, Pr, and Sm) manganites are
able to suppress the robust charge ordered state and effectively
induce ferromagnetic tendencies \cite{S. Hebert,C. Martin,W.
Schuddinck,J.-S. Kang,R. Mahendiran,L. Damari,S. Hebert-1,J. L.
Garcia,B. Raveau,C. Yaicle,Y. Moritomo,T. Kimura,S. Mori,A.
Machida,V. Markovich,R. Mahendiran-1,S. Xu,Z. B. Yan}. Among
these, Ru doping is found to rapidly suppress the
antiferromagnetic phase and stabilize the long range ordered
ferromagnetic metallic phase. Hebert et al. \cite{S. Hebert} have
shown that the elements with $d{}^{0}$ or $d{}^{10}$ configuration
and those without $d$ orbitals does not induce ferromagnetism,
whereas those with $d$ electrons cause suppression of
antiferromagnetic phase at the expense of ferromagnetic state.
However, $Fe^{3+}$ ions despite having $d{}^{5}$ electronic
configuration is shown to belong to the former category,
displaying no evidence of ferromagnetic state in
$Pr{}_{0.5}Ca_{0.5}Mn_{0.95}Fe_{0.05}O_{3}$ sample. In contrast,
the magnetization and transport studies on intermediate bandwidth
$La{}_{0.5}Ca_{0.5}Mn_{1-x}Fe_{x}O_{3}$ compounds show the
favoring of ferromagnetic phase for x < 0.05, while for higher Fe
doping the ferromagnetic transition temperature decreases, leading
the system towards spin glass like state \cite{P. Levy,P.
Levy-1,X. Chen,K. H. Ahn}. Recent investigations show that the
systems with non magnetic dopants ($d{}^{0}$ or $d{}^{10}$) such
as in $Pr{}_{0.5}Ca_{0.5}Mn_{1-x}M_{x}O_{3}$,
$Pr{}_{0.5}Sr_{0.5}Mn_{1-x}M_{x}O_{3}$ (M = Al and Ga) \cite{A.
Banerjee,A. Banerjee-1,S. Nair,A. K. Pramanik}, and
$La{}_{0.5}Ca_{0.5}Mn_{1-x}Ti_{x}O_{3}$, may also favor
ferromagnetic tendencies \cite{W. Tong}. Contrastively, in Ga
substituted $Nd{}_{0.5}Sr_{0.5}MnO_{3}$ system, the gradual
emergence of charge ordered phase at the expense of ferromagnetic
phase is observed \cite{B. Hong}. In
$Pr_{0.5}Ca_{0.5}Mn_{1-x}Al_{x}O_{3}$ \cite{S. Nair} and
$Pr_{0.5}Sr_{0.5}Mn_{1-x}Ga_{x}O_{3}$ \cite{A. K. Pramanik}
series, the contrasting magnetic behavior with Al and Ga doping is
ascribed to preferential substitution of $Mn^{4+}$ with $Al^{3+}$
and $Mn^{3+}$ with $Ga^{3+}$ ions due to their similar ionic
radii. Most of these studies carried out using magnetization and
transport techniques highlight the transition from charge ordered
insulating to ferromagnetic metallic state. The phase separation
behavior as a result of low doping, which has been addressed in
some of the theoretical studies is not clear from these studies.

The neutron diffraction studies in $Pr{}_{0.5}Ca_{0.5}MnO_{3}$ \cite{C. Martin-1}
and $Nd{}_{0.5}Ca_{0.5}MnO_{3}$ \cite{A. Machida} samples have shown
the absence of ferromagnetic state on non magnetic doping of Al ions.
Recently, studies on Co and Ti substituted $Pr{}_{0.5}Ca_{0.5}MnO_{3}$
samples indicate the importance of strain due to the coexistence of
two disimilarly distorted phases, below the charge ordering temperature
$(T_{CO})$ \cite{C. Yaicle,C. Frontera,J. L. Garc=0000EDa-Mu=0000F1oz,C. Frontera-1}.
These two phases present different level of distortions and strains;
the fraction of magnetic phase with large distortion reduces while
one with smaller distortion increases. Neutron diffraction study in
$La{}_{0.5}Ca_{0.5}MnO_{3}$ compound reveals the suppression of orthorhombic
to charge and orbitally ordered monoclinic structural transition upon
doping with Cr or Ni and favors ferromagnetic phase \cite{A. Martinelli}.

The role of B - site substitution on the charge and spin structure
in half doped manganites have also been theoretically investigated
\cite{X. Chen-1,K. Pradhan,K. Pradhan-1,C. Frontera-theo}. The phase
separation behavior observed in these B - site doped compounds has
been attributed to density or defect together with density driven.
These studies also show that the non - magnetic dopants too favor
the stabilization of ferromagnetic state. This behavior is ascribed
to lattice defects and $e_{g}$ carrier density. The destabilization
of charge and orbitally ordered CE - type antiferromagnetic state
is followed by the evolution of ferromagnetic metallic, ferromagnetic
clusters with or without the C - type antiferromagnetic or a spin
glass like phases. Though the previously reported studies have established
the destabilization of charge and orbitally ordered state by non magnetic
and magnetic dopants, very few neutron diffraction studies exist,
which reproduces the theoretically postulated magnetic structures.

In the present work, we have investigated the influence of B - site
disorder on crystal and magnetic structure in $La{}_{0.5}Ca{}_{0.5}MnO_{3}$
system primarily using neutron diffraction and small angle neutron
scattering techniques. The $La{}_{0.5}Ca{}_{0.5}MnO_{3}$ compound
undergoes successive ferromagnetic metallic transition at $T_{C}\approx230K$
and CE - type antiferromagnetic insulating transition at $T{}_{N}\approx170K$
\cite{P. G. Radaelli,P. G. Radaelli-1}. We have studied the polycrystalline
$La{}_{0.5}Ca{}_{0.5}Mn_{1-x}B_{x}O_{3}$ (B = Fe, Ru, Al, and Ga)
compounds. The choice of dopants is due to the contrasting magnetic
ground states these are expected to induce. The destabilization of
charge and orbitally ordered antiferromagnetic state against B - site
disorder is observed. Depending on dopants, these are found to induce
long range ordered ferromagnetic metallic or insulating state, short
range ordered ferromagnetic phase with or without coexisting antiferromagnetic
phases. In compounds with magnetic doping of Fe and Ru, no signature
of magnetic phase coexistence behavior is found. Interestingly, we
have found that non-magnetic substitution (Al or Ga) in $La{}_{0.5}Ca{}_{0.5}MnO_{3}$
compound leads to a coexistence of CE - type state and ferromagnetic
ordering for $x\leq0.03$ and favor A - type antiferromagnetic structure
in $0.05\leq x\leq0.07$ samples. For higher Al and Ga concentration
the system evolves into a magnetically disordered state. This behavior
is as against the magnetic dopants where the evolution of A-type antiferromagnetic
phase is not found. In this study, we also do not observe C - type
antiferromagnetic structure following the melting of CE - type state
shown in the theoretically proposed phase diagram \cite{X. Chen-1}.

\section{Experimental method}

The polycrystalline samples were synthesized by conventional solid-state
reaction method. The starting materials $La{}_{2}O_{3}$, $MnO_{2}$,
$CaCO_{3}$, $Fe{}_{2}O_{3}$, $RuO_{2}$, $Al{}_{2}O_{3}$, and $Ga{}_{2}O_{3}$
were mixed in proper stoichiometric ratio and fired at 1250$^{\circ}C$
for 48 hrs. Samples were then repeatedly ground and heat treated at
1400$^{\circ}C$ for 48 hrs. Finally, samples were pelletized and
sintered at 1450$^{\circ}C$ for 48 hrs. Initial values of cell parameters
and phase identification of all the samples at 300K were obtained
from X-ray powder diffraction pattern recorded on a Rigaku diffractometer,
rotating anode type using Cu K$\alpha$ radiation. Neutron diffraction
patterns were recorded on a multi PSD based powder diffractometer
($\lambda$ = 1.249 and 1.2443$\text{\AA}$) at Dhruva reactor, Bhabha
Atomic Research Centre, Mumbai, at selected temperatures between 5
and 300K, in the $5^{\circ}\leq2\theta\leq140^{\circ}$ angular range.
The powdered samples were packed in a cylindrical Vanadium container
and attached to the cold finger of a closed cycle Helium refrigerator.
Rietveld refinement of the diffraction patterns were performed using
FULLPROF program \cite{J. Rodriguez-Carvajal}. The magnetization
measurements were carried out using SQUID (Quantum design, USA) /
or VSM magnetometer (oxford instruments). Standard four point probe
technique was used to measure the dc resistivity between 3 and 300K.
Small angle neutron scattering (SANS) measurements $(\lambda=4.52$
and $7.03\text{\AA)}$ as a function of temperature between 15 and
300K in the Q range $0.007\leq Q\leq0.3\lyxmathsym{\AA}^{\text{\textminus}1}$
were carried out on the SANS - II instrument at the Swiss Spallation
Neutron Source, Paul Scherrer Institute, Switzerland \cite{P. Strunz}.

\section{Results and discussion}

\subsection{Crystal Structure}

All the B - site doped compounds studied here in the series $La{}_{0.5}Ca{}_{0.5}Mn_{1-x}B_{x}O_{3}$
(B = Fe $(0.01\leq x\leq0.10)$, Ru $(0.01\leq x\leq0.05)$, Al $(0.01\leq x\leq0.13)$,
and Ga $(0.01\leq x\leq0.07)$) are isostructural, possessing orthorhombic
structure (space group\textit{ Pnma}) at 300K. These compounds crystallize
in $O{}^{\prime}$ orthorhombic symmetry characterized by $b/\surd2\leq a\leq c$
\cite{B. B. van Aken,G Maris}. The refinement of x - ray diffraction
patterns at 300K reveal that volume as a function of doping (x) with
B = Fe, Ru, Al, and Ga do not change appreciably and show any systematic
behavior. This could be due to small concentration of dopants and
very small difference in the value of ionic radii between Mn ions
and the dopants \cite{R. D. Shannon}.

\noindent %
\begin{figure}[t]
\resizebox{0.5\textwidth}{!}{
\includegraphics{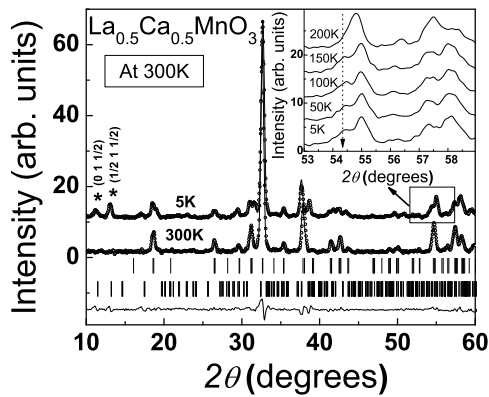}}

\caption{\label{fig:ND_LCMO}A neutron diffraction pattern of $La{}_{0.5}Ca_{0.5}MnO{}_{3}$
sample at 5 and 300K. The symbol ({*}) indicates the CE-type antiferromagnetic
superlattice reflections. Continuous lines through the data points
are the fitted lines to chemical and magnetic structures described
in the text. The vertical tick marks below the pattern correspond
to indexing of nuclear and magnetic peaks. Curve at the bottom is
the difference between observed and calculated intensity at 5K. The
inset to the figure shows a selected portion $(53\lyxmathsym{\textdegree}\leq2\theta\leq58.5\lyxmathsym{\textdegree})$
of neutron diffraction patterns at various temperatures for $La{}_{0.5}Ca_{0.5}MnO{}_{3}$
sample. The arrow indicates splitting of (0 0 4) (4 0 0) (2 4 2) peak
in \textit{Pnma} space group to (0 0 8) (4 4 2) and ($\overline{4}$
4 2) in $P2{}_{1}/m$ space group, marking the orthorhombic to monoclinic
structural transition.}

\end{figure}

Figure \ref{fig:ND_LCMO} shows the neutron diffraction pattern of
$La{}_{0.5}Ca{}_{0.5}MnO_{3}$ (x = 0) compound. On lowering of
temperature below the charge ordering temperature,
$T_{CO}\approx200K$ (on heating), structural transition to a lower
symmetry monoclinic structure in $P2_{1}/m$ space group is found,
as has been reported earlier in similar systems exhibiting the
charge ordering behavior \cite{P. G. Radaelli}. A signature of
this in the form of splitting of nuclear Bragg reflections (0 0 4)
(4 0 0) (2 4 2) in \textit{Pnma} space group to (0 0 8) (4 4 2)
and ($\overline{4}$ 4 2) in $P2{}_{1}/m$ space group is observed
below $T_{CO}$. The patterns displayed in the inset to figure
\ref{fig:ND_LCMO} shows the structural transformation of
orthorhombic phase in \textit{Pnma} space group to charge and
orbitally ordered monoclinic phase in $P2_{1}/m$ space group. In
$P2{}_{1}/m$ space group, $Mn^{3+}$ and $Mn^{4+}$ ions occupy two
distinct sites, in accordance with charge order scenario proposed
in the Goodenough model \cite{J. B. Goodenough-1}. Nevertheless,
the Goodenough model has been challenged by experiments, which
have presented evidence for charge disproportion in several half
doped manganites. Studies reveal that this ionic picture proposed
in Goodenough model of $Mn^{3+}$ and $Mn^{4+}$ ions may not be
true and Zener polaron model \cite{A. Daoud-Aladine,L. Wu} has
been proposed, in which all the Mn sites become equivalent with a
valence of +3.5. Herrero-Martin et al. have probed the local
structure around Mn ions in charge ordered systems using resonant
scattering of synchrotron x-ray beam and show the presence of
fractional charge segregation \cite{J. Herrero-Martin}. This
disagreement between the two models is still not settled in
literature. We find, in the absence of high resolution data,
similar values of $\chi^{2}$ and R-factors on fitting the
diffraction data in both \textit{Pnma} and $P2{}_{1}/m$ space
groups. Additionally, lowering of symmetry to $P2{}_{1}/m$ space
group requires the refinement of 29 positional parameters as
against 7 in \textit{Pnma} space group. This reduces the
reliability of obtained positional parameters. Therefore, the low
temperature crystal structure is refined in \textit{Pnma} space
group, which yields an average structure. Similar refinement of
neutron diffraction pattern has been performed in
$La{}_{0.5}Ca{}_{0.5}MnO_{3}$ compound, which is in agreement with
Radaelli et al. \cite{P. G. Radaelli}. The obtained values of
structural parameters are in agreement with the previously
reported studies on $La{}_{0.5}Ca{}_{0.5}MnO_{3}$ compound
\cite{P. G. Radaelli}.

\noindent %
\begin{figure}[t]
\resizebox{0.5\textwidth}{!}{
\includegraphics{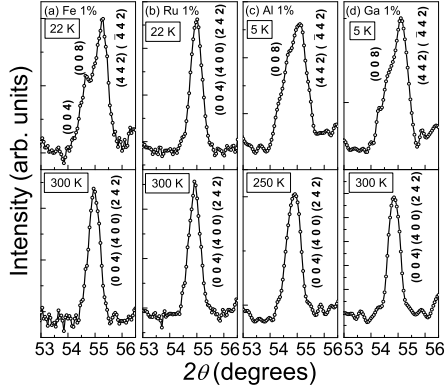}}

\caption{\label{fig:ND_selected}A selected portion $(53\lyxmathsym{\textdegree}\leq2\theta\leq56.5\lyxmathsym{\textdegree})$
of neutron diffraction patterns collected around the orthorhombic
to monoclinic structural transition for $La{}_{0.5}Ca_{0.5}Mn_{0.99}B_{0.01}O{}_{3}$
samples in (a) B = Fe, (b) B = Ru, (c) B = Al, and (d) B = Ga.}

\end{figure}

For comparison the selected portion of neutron powder diffraction
patterns between $53^{\circ}\leq2\theta\leq56.5{}^{\circ}$ for $La{}_{0.5}Ca{}_{0.5}Mn_{0.99}B_{0.01}O_{3}$
(B = Fe, Ru, Al, and Ga) samples is shown in figure \ref{fig:ND_selected}.
In $La{}_{0.5}Ca{}_{0.5}Mn_{0.99}Fe_{0.01}O_{3}$ (Fe 1\% doping)
sample the orthorhombic to monoclinic structural transition (figure
\ref{fig:ND_selected}(a)), similar to x = 0 compound is observed.
In samples with higher Fe doping $(0.02\leq x\leq0.10)$, this splitting
is suppressed. This indicates the suppression of charge and orbital
ordering with increasing Fe doping. In contrast, the similar doping
concentration of Ru (1\%) is found to suppress the monoclinic structural
transformation and the orthorhombic structure in \textit{Pnma }space
group is retained down to the lowest temperature of 22K, as shown
in figure \ref{fig:ND_selected}(b). On substitution with Al and Ga
$(x\leq0.03)$, the signature of structural transformation from orthorhombic
to charge and orbitally ordered monoclinic structure is observed,
as shown in figure \ref{fig:ND_selected}(c) and (d). On increasing
the doping level in $x\geq0.03$ samples, the signature for the presence
of monoclinic structure gradually disappears. This behavior is similar
to that observed in Ni and Cr doped $La{}_{0.5}Ca{}_{0.5}MnO_{3}$
compounds by Martinelli et al. using neutron diffraction technique\cite{A. Martinelli}.

\noindent %
\begin{figure}[t]
\resizebox{0.5\textwidth}{!}{
\includegraphics{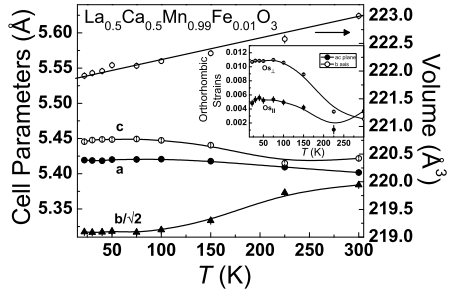}}

\caption{\label{fig:Cell_Fe01}Temperature dependence of lattice parameters
and the unit-cell volume for $La{}_{0.5}Ca_{0.5}Mn_{0.99}Fe_{0.01}O{}_{3}$
(Fe 1\%) sample. The inset to the figure shows the temperature dependence
of orthorhombic strains $Os_{\bot}$ and $Os_{\parallel}$ in $La{}_{0.5}Ca_{0.5}Mn_{0.99}Fe_{0.01}O{}_{3}$
sample. The continuous lines are a guide for the eye. }

\end{figure}

\begin{figure}[t]
\resizebox{0.5\textwidth}{!}{
\includegraphics{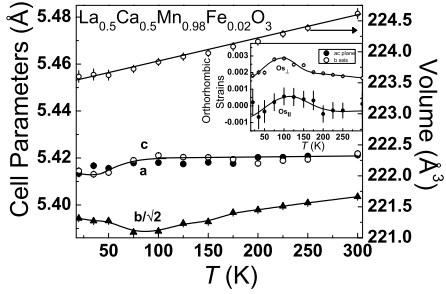}}

\caption{\label{fig:Cell_Fe02}Temperature dependence of lattice parameters
and the unit-cell volume for $La{}_{0.5}Ca_{0.5}Mn_{0.98}Fe_{0.02}O{}_{3}$
(Fe 2\%) sample. The inset to the figure shows the temperature dependence
of orthorhombic strains $Os_{\bot}$ and $Os_{\parallel}$ in $La{}_{0.5}Ca_{0.5}Mn_{0.98}Fe_{0.02}O{}_{3}$
sample. The continuous lines are a guide for the eye. }

\end{figure}

The cell parameters as a function of temperature in 1\% Fe doped sample
is shown in figure \ref{fig:Cell_Fe01}. On reducing the temperature
below 300K, lattice parameters exhibit an anomalous behavior, while
volume displays nearly linear temperature dependence. The lattice
parameter b shows a drastic decrease while a and c expand on lowering
of temperature. As a consequence, in neutron diffraction pattern the
(2 0 2) and (0 4 0) nuclear Bragg reflections which were merged at
300K, exhibit splitting while cooling below $T\leq225K$ in 1\% Fe
doped compound. This anomalous behavior is associated with ordering
of $d{}_{z^{2}}$ orbitals in the a-c plane, which is in agreement
with other similar charge ordered systems \cite{P. G. Radaelli}.
The temperature at which orbital ordering occurs is found to coincide
with the charge ordering temperature. The charge ordering transition
is the temperature below which the resistivity shows a steep rise.
The transition to charge and orbitally ordered state is accompanied
with an increase of strain in the a-c plane, as evidenced in $La{}_{0.5}Ca{}_{0.5}MnO_{3}$
compound \cite{P. G. Radaelli-1}. Also, Ahn et al. have reported
the importance of uniform strain in stabilization of the charge ordered
phase \cite{K. H. Ahn_strain}. To describe the orthorhombic strains,
Meneghini et al. defined $Os{}_{\shortparallel}=2\left(\frac{c-a}{c+a}\right)$
distortions in the ac plane and $Os{}_{\bot}=2\left(\frac{c+a-b\sqrt{2}}{c+a+b\sqrt{2}}\right)$
along the b axis \cite{C. Meneghini}. The charge and orbital ordering
transition is accompanied with a sharp increase in the orthorhombic
strain parameters, $Os{}_{\shortparallel}$ and $Os{}_{\bot}$ \cite{P. G. Radaelli-1}.
In 1\% Fe doped compound, on lowering of temperature to 225K, both
$Os{}_{\shortparallel}$ and $Os{}_{\bot}$ increase and become nearly
constant below 100K, as shown in the inset to figure \ref{fig:Cell_Fe01}.
The strain increases below $T_{CO}$ and therefore, favors the stabilization
of charge ordered phase. Figure \ref{fig:Cell_Fe02} shows the variation
of cell parameters as a function of temperature for 2\% Fe doped compound.
In contrast to 1\% Fe doped sample, here the cell parameters a, b
and c decrease on lowering of temperature. However, below 100K, an
anomalous behavior is observed. A minimum in b is found at 100K, which
coincides with the onset of ferromagnetic transition temperature $(T{}_{C})$.
This plausibly indicates that the onset of ferromagnetic ordering
is accompanied by the reorientation of orbitals. The temperature dependence
of lattice parameters in 4\% Fe doped samples exhibit similar decrease
on lowering of temperature and minimum in lattice parameter b at 125K,
coinciding with $T_{C}$. Similar temperature dependence of lattice
parameters, with a minimum in b close to $T_{C}$, has been reported
previously in neutron diffraction studies on $La{}_{0.5}Ca{}_{0.5}Mn_{1-x}B{}_{x}O_{3}$,
for B = Ni and Cr doped systems \cite{A. Martinelli}. The orthorhombic
strains $Os{}_{\shortparallel}$ and $Os{}_{\bot}$, for 2\% (shown
in the inset to figure \ref{fig:Cell_Fe02}) and 4\% Fe doped samples,
exhibit a maximum close to ferromagnetic transition temperature of
100 and 125K, respectively. Such a behavior may indicate that with
the onset of long range ordered ferromagnetic phase below $T_{C}$,
orthorhombic strains in the lattice are reduced, accompanied with
suppression of charge and orbital ordering. At higher Fe doping in
$La{}_{0.5}Ca{}_{0.5}Mn_{1-x}Fe_{x}O_{3}$ with $x\geq0.06$, the
cell parameters exhibit nearly linear reduction while cooling. As
a consequence, no significant change in $Os{}_{\shortparallel}$ and
$Os{}_{\bot}$ is observed.

\noindent %

In $La{}_{0.5}Ca{}_{0.5}Mn_{1-x}Ru{}_{x}O_{3}$ with $0.01\leq x\leq0.03$
samples, the cell parameters as a function of temperature display
behavior similar to that described for 2\% Fe doped sample. A minimum
in lattice parameter b is observed at $\sim200K$, coinciding with
the ferromagnetic transition temperature. In accompaniment, the $Os{}_{\shortparallel}$
and $Os{}_{\bot}$ show a maximum at temperature close to $T_{C}$.
On increasing the Ru doping in $La{}_{0.5}Ca{}_{0.5}Mn_{0.95}Ru_{0.05}O_{3}$
(Ru 5\% doping) sample, no anomalous behavior in lattice parameters
as a function of temperature is observed and display almost linear
reduction on lowering of temperature below 300K. Consequently, the
orthorhombic strain parameters exhibit no significant change as a
function of temperature.

With Al doping in $La{}_{0.5}Ca{}_{0.5}Mn_{1-x}Al_{x}O_{3}$ $(0.01\leq x\leq0.07)$
compounds, the temperature evolution of cell parameters exhibit behavior
typical of charge and orbital ordering, as observed in $La{}_{0.5}Ca{}_{0.5}Mn_{0.99}Fe_{0.01}O_{3}$
sample (figure \ref{fig:Cell_Fe01}). The variation of orthorhombic
strain parameters show an increase below the charge and orbital ordering
temperature. This indicates the ordering of $d_{z^{2}}$ orbitals
in the ac plane is accompanied with an increase in strain \cite{P. G. Radaelli,P. G. Radaelli-1}.
Additionally, in the 3\% Al doped sample, broad minimum in lattice parameter
b accompanied by a maximum in orthorhombic strains close to ferromagnetic
transition temperature $(T_{C}\approx125K)$ is observed. In contrast
with Fe and Ru substituted $La_{0.5}Ca_{0.5}MnO_{3}$ compounds, the
signature of charge and orbital ordering in Al doped systems are retained
up to much higher doping level with x = 0.07. However, the transition
temperature of the charge and orbital ordering continuously reduces
with increasing Al doping. Finally, in $0.10\leq x\leq0.13$ samples,
charge and orbitally ordered state is fully suppressed. Similarly,
with Ga doping of 1\%, the signature of charge and orbital ordering
is evident in the temperature dependence of cell parameters. At 3\%
Ga doping, minimum in lattice parameters is seen at $T\approx125K$.
This behavior is similar to 2 - 4\% Fe and 1 - 3\% Ru doped samples.
At higher Ga substitution, charge and orbitally ordered state is fully
suppressed.

Therefore, doping with magnetic ions (Fe and Ru) causes suppression
of charge and orbital ordering much more rapidly than with non magnetic
dopants (Al and Ga). In samples exhibiting ferromagnetic ordering,
anomalous behavior of cell parameters is observed, which coincides
with the onset of ferromagnetic transition. These results indicate
that with induced disorder at B - site the homogeneous strain field
is collapsed into an inhomogeneous one, which is accompanied by the
gradual suppression of charge ordered phase. Theoretical studies have
shown that long range homogeneous strain plays a crucial role in stabilization
of charge and orbitally ordered state in half doped manganites \cite{J. Burgy}.
Experimentally, the presence of phase coexistence behavior in manganites
is correlated with different lattice strains, wherein their interplay
with temperature lead to stabilization of one phase at the expense
of the other \cite{C. Yaicle,C. Frontera,J. L. Garc=0000EDa-Mu=0000F1oz,C. Frontera-1}.
Structurally, more distorted phase would favor insulating state, whereas
the less distorted one would exhibit ferromagnetic metallic behavior.

\subsection{Magnetization and Transport Behavior}

\noindent %
\begin{figure}[t]
\resizebox{0.5\textwidth}{!}{
\includegraphics{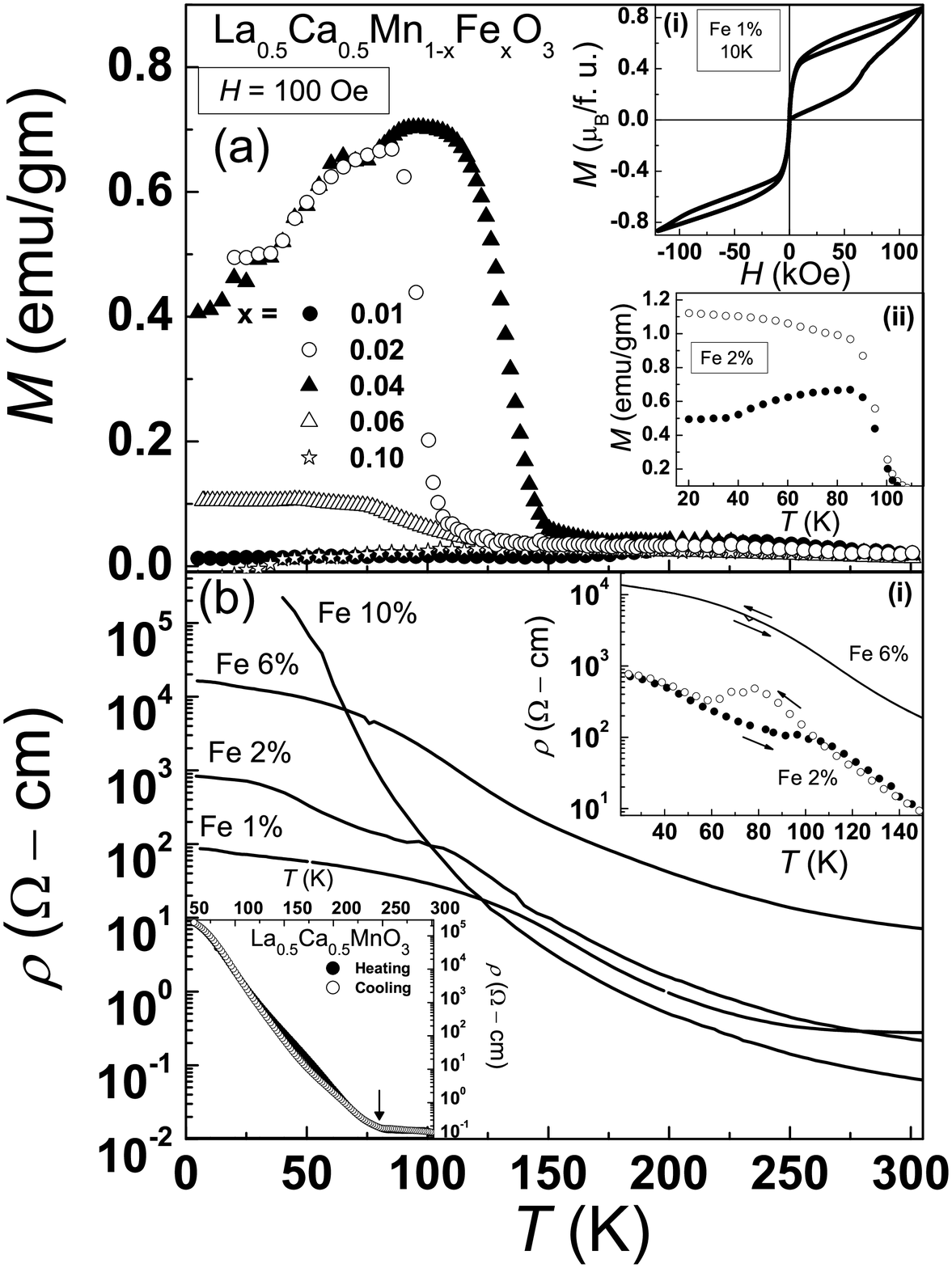}}

\caption{\label{fig:M(T)_R(T)_Fe}(a) Magnetization and (b) resistivity as
a function of temperature for $La{}_{0.5}Ca_{0.5}Mn_{1-x}Fe_{x}O{}_{3}$
$(0.01\leq x\leq0.10)$ samples. Inset (i) to figure (a) shows variation
of magnetization (M) with field (H) for 1 \% Fe doped sample and inset
(ii) shows the ZFC (closed symbols) and FC (open symbols) magnetization,
displaying the bifurcation behavior for 2\% Fe doped sample. While
the inset (i) to figure (b) displays a thermal hysteresis in $\rho(T)$
between heating (closed symbols) and cooling (open symbols) cycles
in $La{}_{0.5}Ca_{0.5}Mn_{1-x}Fe_{x}O{}_{3}$ samples with x = 0.02
and 0.06. From this figure, the gradual reduction in the width of
thermal hysteresis with increasing Fe doping is evident. The inset
(ii) to figure (b) shows the temperature dependence of resistivity
for heating (closed symbols) and cooling (open symbols) cycles in
$La{}_{0.5}Ca_{0.5}MnO{}_{3}$ (x = 0) sample and arrow indicates
the transition to a charge ordered state $(T_{CO})$, below which
resistivity shows a steep rise.}

\end{figure}

The temperature dependence of magnetization, M(T), measured under
zero field cooled (ZFC) and field cooled (FC) conditions for $La{}_{0.5}Ca{}_{0.5}Mn_{1-x}Fe_{x}O_{3}$
$(0.01\leq x\leq0.10)$ samples, is shown in figure \ref{fig:M(T)_R(T)_Fe}(a).
The 1\% Fe doped sample undergoes multiple magnetic transitions as
a function of temperature, displaying behavior similar to x = 0 compound
\cite{P. G. Radaelli-1}. On lowering of temperature at $\sim230K$,
a hump in magnetization is observed (not visible on the present scale
in figure \ref{fig:M(T)_R(T)_Fe}(a)). This coincides with the onset
of charge ordered insulating state and $T_{C}$, as observed for x
= 0 compound \cite{P. G. Radaelli-1}. On further reducing the temperature,
another broad hump in magnetization is observed at $\sim100K$, which
is identified with an antiferromagnetic transition temperature $(T_{N})$
obtained from neutron diffraction study. The variation of magnetization
with field, M(H), of 1\% Fe doped sample at T = 10K is shown in the
inset to figure \ref{fig:M(T)_R(T)_Fe}(a). In the region $T<T_{N}$,
coexistence of ferromagnetic clusters in the charge and orbitally
ordered antiferromagnetic matrix is evident from the narrow hysteresis
loop in M(H) behavior. However, the moment value is considerably lower
than the expected value. In addition, anomalous behavior in the form
of virgin curve lying outside the envelope curve and a step like behavior
in virgin curve is visible in M(H) data. Previous reports on the high
resolution neutron diffraction studies in the presence of magnetic
field show that the steps in M(H) of charge ordered systems with CE-type
antiferromagnetic structure is accompanied by a change in the cell
parameters \cite{C. Yaicle-1,V. Hardy}. In Ga substituted $Pr{}_{0.5}Ca_{0.5}Mn_{0.97}Ga_{0.03}O{}_{3}$
compound, magnetic field as high as 6T is observed to favor ferromagnetic
phase, without significantly influencing the CE-type antiferromagnetic
state. A similar description based on martensitic like scenario on
the strain accommodation in phase separated manganites has been proposed
to explain the step like behavior in M(H) \cite{V. Hardy,V. Hardy-1}.
At higher Fe doping in $La_{0.5}Ca_{0.5}Mn_{1-x}Fe_{x}O_{3}$ with
$0.02\leq x\leq0.06$, the magnetic nature appears to be modified.
On cooling below 300K, the M(T) exhibits an increase below the respective
ordering temperatures and display no significant change thereafter,
which indicates the onset of ferromagnetic state. The ferromagnetic
transition temperature, $T_{C}$, for 2\%, 4\%, and 6\% Fe doped samples
are 100, 150 and 75K, respectively. In these Fe doped samples, a large
bifurcation between the ZFC and FC curves arises below 50K, as shown
in the inset (ii) to figure \ref{fig:M(T)_R(T)_Fe}(a) for 2\% Fe
doped sample, indicating the coexistence of ferromagnetic and spin
glass phase. The presence of spin glass phase in B - site substituted
compounds has been brought out in some of the recent theoretical studies
\cite{C. Frontera-theo}. The ferromagnetic behavior is progressively
suppressed with increase in Fe doping and in 10\% Fe doped sample,
no evidence of long range magnetic ordering is observed. This is also
confirmed by neutron diffraction measurements described later. However,
extremely narrow hysteresis loop evident in M(H) suggests the presence
of short range ferromagnetic ordering.

The variation of resistivity with temperature, $\rho(T)$, in $La{}_{0.5}Ca{}_{0.5}Mn_{1-x}Fe_{x}O_{3}$
$(0.01\leq x\leq0.10)$ samples is shown in figure \ref{fig:M(T)_R(T)_Fe}(b).
The resistivity data were collected during heating and cooling cycles.
In these samples, the temperature dependence of resistivity exhibits
an insulating behavior over the entire measured temperature range
between 5 to 300K. In comparison to $La{}_{0.5}Ca{}_{0.5}MnO_{3}$
compound ($\rho(T)$ shown in the inset (ii) to figure \ref{fig:M(T)_R(T)_Fe}(b)),
in Fe doped samples with $0.01\leq x\leq0.06$, reduction in resistivity
at 5K by nearly two orders of magnitude is found. Besides, with Fe
doping the point of inflection $(\approx T_{CO})$ below which the
$\rho(T)$ exhibits a steep rise is absent. This indicates the suppression
of charge ordering in these samples, which is in agreement with the
suppression of structural transition from orthorhombic phase in \textit{Pnma}
space group to monoclinic phase in $P2_{1}/m$ space group observed
in the low temperature neutron diffraction study. Additionally, a
broad hump accompanied with a thermal hysteresis between heating and
cooling cycle, is evident near $T_{C}$ and is shown in the inset
(i) to figure \ref{fig:M(T)_R(T)_Fe}(b). The gradual suppression
of the anomalous behavior and reduction in the width of thermal hysteresis
with increasing Fe doping is also evident in this figure. The thermal
hysteresis behavior in $\rho(T)$, observed in phase separated manganites,
has been explained to arise from the formation of metal clusters and
their size distribution across the metal insulator transition \cite{D. Khomskii}.

\noindent %
\begin{figure}[t]
\resizebox{0.5\textwidth}{!}{
\includegraphics{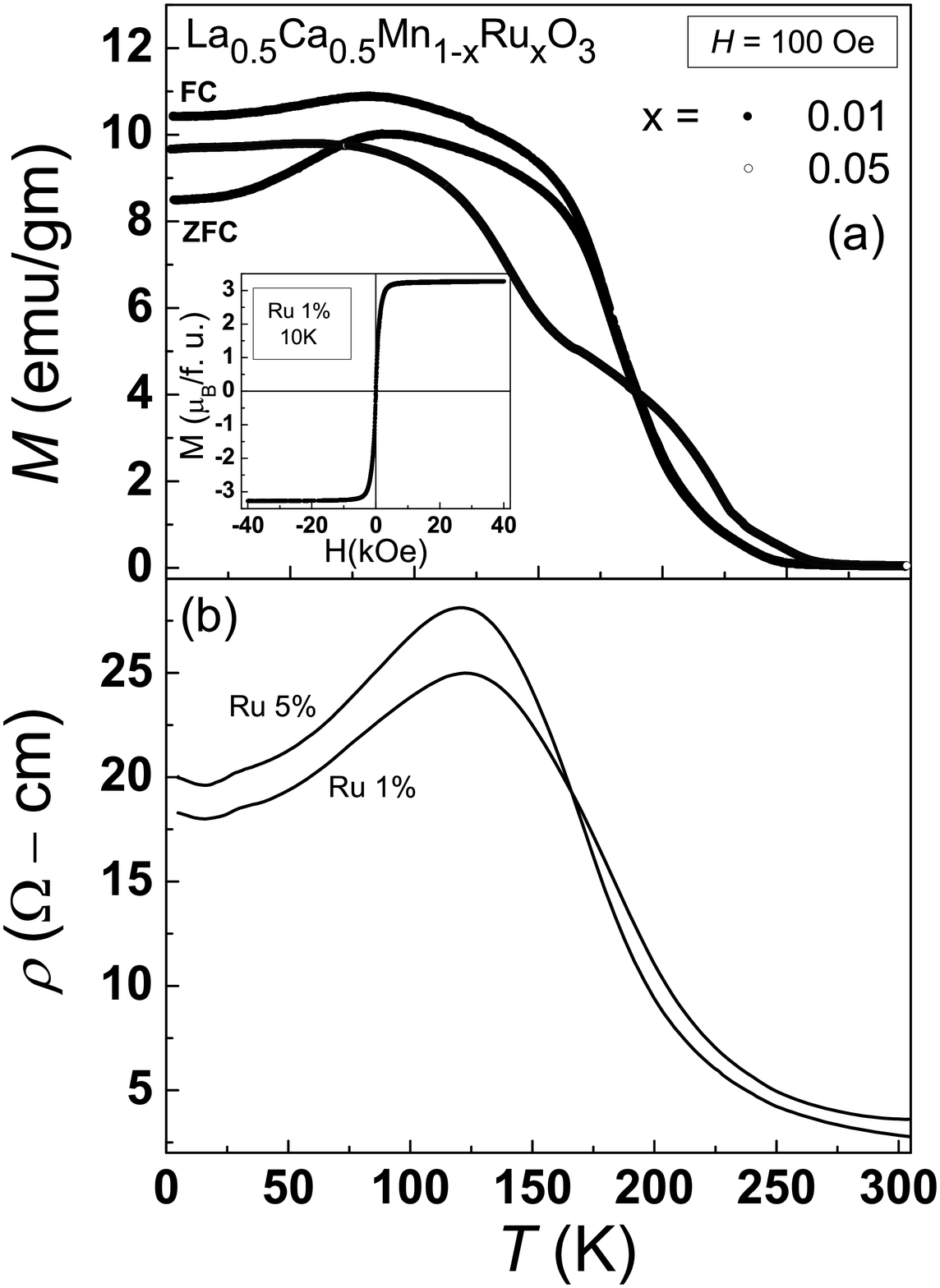}}

\caption{\label{fig:M(T)_R(T)_Ru}(a) Magnetization as a function of temperature
for $La{}_{0.5}Ca_{0.5}Mn_{1-x}Ru_{x}O{}_{3}$ $(0.01\leq x\leq0.05)$
samples and the inset shows variation of magnetization (M) with field
(H) in $La{}_{0.5}Ca_{0.5}Mn_{0.99}Ru{}_{0.01}O{}_{3}$ (Ru 1\%) sample.
In (b) variation of resistivity with temperature is shown. }

\end{figure}

The magnetization data, M(T), of $La{}_{0.5}Ca{}_{0.5}Mn_{1-x}Ru{}_{x}O_{3}$
$(0.01\leq x\leq0.05)$ samples is shown in figure \ref{fig:M(T)_R(T)_Ru}(a).
As a function of Ru doping, the increase in M(T) at 200K $(\sim T_{C})$
indicates an onset of ferromagnetic ordering. The broad nature of
transition indicates an inhomogeneous nature of ferromagnetic state.
The M(H) data, for 1\% Ru doped sample is shown in the inset to figure
\ref{fig:M(T)_R(T)_Ru}(a). This figure is representative of the Ru
doped samples with $0.01\leq x\leq0.05$. The signature of a typical
ferromagnetic behavior is evident in M(H) curve. At low field, magnetization
M shows a sharp rise and saturates thereafter. The saturation value
of magnetization $(\sim3.3\mu{}_{B}/f.u.)$ is close to the expected
value of $\sim3.5\mu{}_{B}/f.u.$ In figure \ref{fig:M(T)_R(T)_Ru}(b),
temperature dependence of resistivity, $\rho(T)$, of Ru doped $La_{0.5}Ca_{0.5}Mn_{1-x}Ru_{x}O_{3}$
$(0.01\leq x\leq0.05)$ samples is displayed. With Ru doping of 1\%
the charge ordered insulating state is destabilized and insulator
to metal transition is induced at $T_{MI}\approx125K$. Similar behavior
in $\rho(T)$ is observed for Ru doped sample with x = 0.05. No significant
change in $T_{C}$ and $T_{MI}$ as a function of Ru doping is observed.
Similar behavior has been recently reported in Ru doped $Sm_{0.55}Sr_{0.45}MnO_{3}$
compounds, attributed to the coexistence and competing nature of double
exchange and superexchange interactions \cite{Y. Ying,M. M. Saber}.
Also, it is observed that there is a large difference between $T_{C}$
and $T_{MI}$ in Ru substituted samples; $\triangle T=T_{C}-T_{MI}=75K$.
The observed difference indicates the percolative nature of ferromagnetic
regions.

\noindent %
\begin{figure}[t]
\resizebox{0.5\textwidth}{!}{
\includegraphics{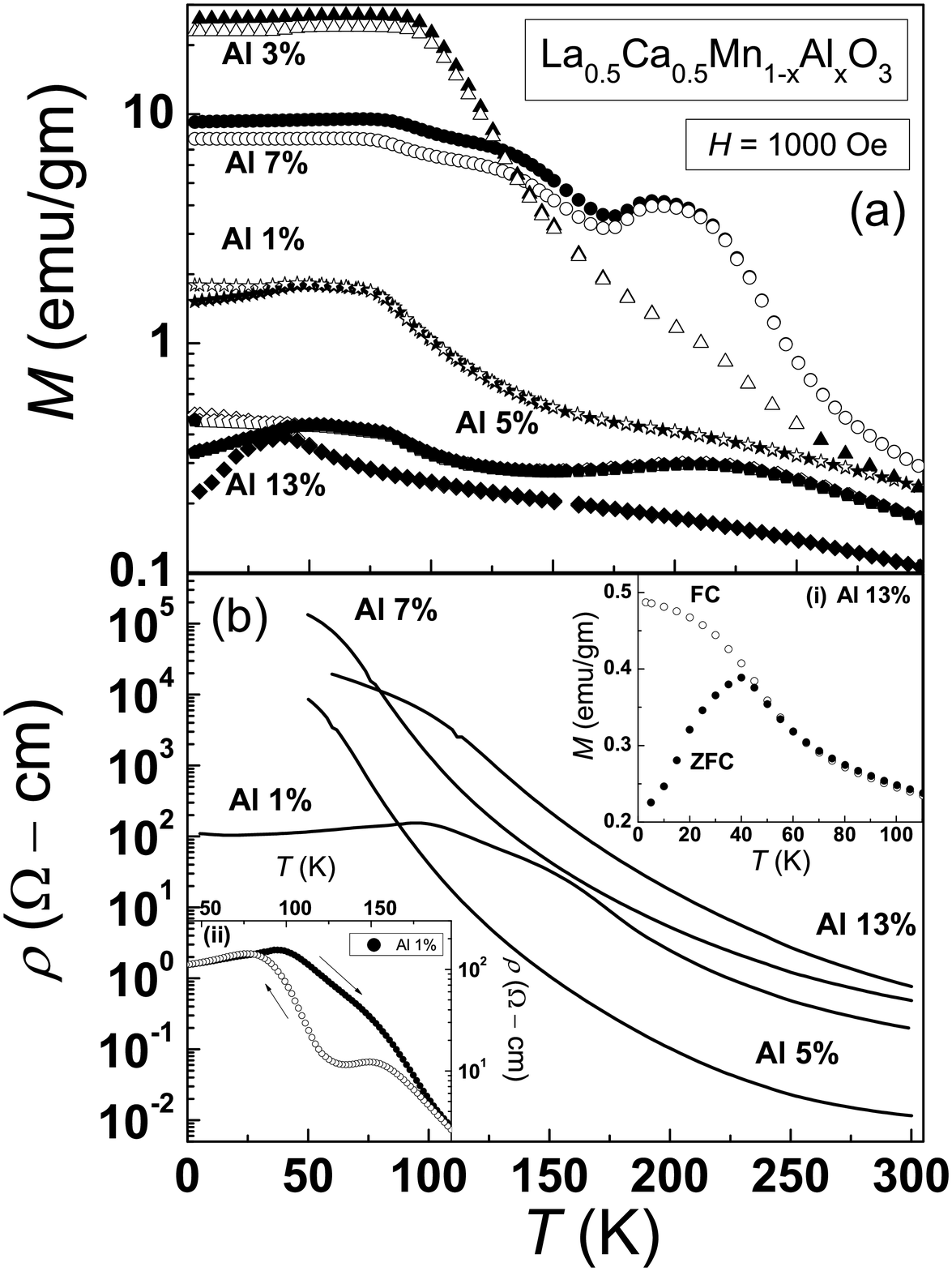}}

\caption{\label{fig:M(T)_R(T)_Al}The temperature dependence of (a)
ZFC (closed symbols) and FC (open symbols) magnetization and (b)
resistivity for $La{}_{0.5}Ca_{0.5}Mn_{1-x}Al_{x}O{}_{3}$
$(0.01\leq x\leq0.13)$ samples. The inset (i) to figure (b) shows
the ZFC (closed symbols) and FC (open symbols) magnetization data
of $La{}_{0.5}Ca_{0.5}Mn_{0.87}Al_{0.13}O{}_{3}$ (Al 13\%) sample
in the temperature range between 5K to 100K and (ii) displays a
thermal hysteresis in $\rho(T)$ between heating (closed symbols)
and cooling (open symbols) cycles for 1\% Al doped sample.}

\end{figure}

In figure \ref{fig:M(T)_R(T)_Al}(a) the magnetization as a
function of temperature in
$La{}_{0.5}Ca{}_{0.5}Mn_{1-x}Al{}_{x}O_{3}$ $(0.01\leq x\leq0.13)$
samples is displayed. In the 1\% Al doped sample, a shallow hump in
M(T) at $\approx200K$ is observed. The SANS study described later
show the onset of short range ordered ferromagnetic correlations
below 200K. On further lowering of temperature the magnetization
saturates below 75K. From the magnetization studies transition to
an antiferromagnetic state is not clear. Neutron diffraction study
on 1\% Al doped sample show evidence of antiferromagnetic ordering
with $T_{N}\approx150K$. The resulting competing nature of the
ferromagnetic and antiferromagnetic interactions describes the
magnetization behavior. The value of magnetization at 5K increases
with increase in Al concentration, exhibiting a maximum for 3\% Al
doping, and decreases subsequently for higher Al doping. In 3\% Al
doped sample transition to a ferromagnetic state occurs at
$T_{C}\approx200K$ and from neutron diffraction study we learn the
presence of antiferromagnetic phase below $T_{N}\approx120K$. In
$La{}_{0.5}Ca{}_{0.5}Mn_{1-x}Al{}_{x}O_{3}$ with $0.05\leq
x\leq0.07$ compounds, similar behavior of transition to
antiferromagnetic and ferromagnetic states are observed. The
nature of antiferromagnetic ordering in these compounds obtained
from neutron diffraction measurements is discussed below.
Likewise, in Al doped samples with $0.10\leq x\leq0.13$,
M(T)increases below 100K and a maximum at 50K is observed. At 50K,
M(T) displays a drop accompanied by bifurcation in FC and ZFC
curves, as shown in the inset (i) to figure
\ref{fig:M(T)_R(T)_Al}(b) in 13\% Al doped sample. Such a behavior
has been reported in other charge ordered manganites and is
ascribed to transition to a spin glass like state \cite{H. Y.
Hwang,J. A. Mydosh}. Both randomness and frustration of spins are
necessary to produce a spin glass state. The competing
antiferromagnetic and ferromagnetic interactions could lead to
such frustrations in the system, leading to a growth of spin glass
or clusters glass state. The nature of spin glass phases in
manganites is still a topic of discussion \cite{J. A. Mydosh,E.
Dagotto-2}. The resistivity data, $\rho(T)$, of Al doped samples
in $La_{0.5}Ca_{0.5}Mn_{1-x}Al_{x}O_{3}$ $(0.01\leq x\leq0.13)$ is
displayed in figure \ref{fig:M(T)_R(T)_Al}(b). In
$La_{0.5}Ca_{0.5}Mn_{1-x}Al_{x}O_{3}$ with $(0.01\leq x\leq0.03)$
samples, while cooling from 300K insulator to metal transition,
accompanied by thermal hysteresis (inset (ii) to figure
\ref{fig:M(T)_R(T)_Al}(b)), is observed at $T_{MI}\approx100K$.
The origin of thermal hysteresis is similar to that described in
the case of Fe doped samples. In higher Al doped compounds with
$0.05\leq x\leq0.13$, temperature dependence of resistivity again
shows an insulating behavior over the entire measured temperature
range between 50 and 300K. Below 50K resistivity is too high to be
measurable. Also, thermal hysteresis between heating and cooling
cycle is diminished.

\noindent %
\begin{figure}[t]
\resizebox{0.5\textwidth}{!}{
\includegraphics{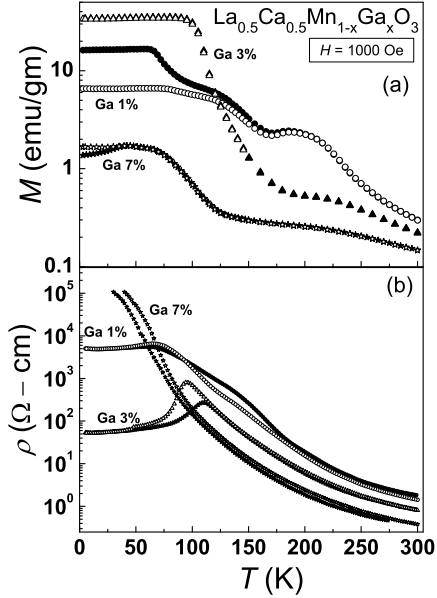}}

\caption{\label{fig:M(T)_R(T)_Ga}(a) ZFC (closed symbols) and FC (open symbols)
magnetization and (b) resistivity as a function of temperature while
heating (closed symbols) and cooling (open symbols) for $La{}_{0.5}Ca_{0.5}Mn_{1-x}Ga{}_{x}O{}_{3}$
$(0.01\leq x\leq0.07)$ samples.}

\end{figure}

In $La{}_{0.5}Ca{}_{0.5}Mn_{1-x}Ga{}_{x}O_{3}$ $(0.01\leq x\leq0.07)$,
the M(T) shown in figure \ref{fig:M(T)_R(T)_Ga}(a), displays behavior
similar to Al doped samples. In 1\% Ga doped sample, the transition
correlated with charge and orbitally ordered state at $\sim180K$
and the magnetic transitions at $T_{N}\approx150K$ (obtained from
neutron diffraction study) and $T_{C}\approx60K$ are observed. At
3\% Ga doping, the ferromagnetic transition is evident at $\sim125K$,
with no bifurcation in ZFC and FC curves. At higher Ga doping of 7\%,
sharp increase in magnetization occurs below $100K$, exhibiting a
maximum at $50K$. Below $50K$, the sharp drop in magnetization is
accompanied with a bifurcation between ZFC and FC curve. Also, the
typical features of metamagnetic transition and phase coexistence
are evident in the M(H) curves of Ga doped samples. With Ga doping
in $La_{0.5}Ca_{0.5}Mn_{1-x}Ga_{x}O_{3}$ $(0.01\leq x\leq0.07)$,
resistivity as a function of temperature display behavior similar
to Al doped sample, as observed in figure \ref{fig:M(T)_R(T)_Ga}(b).
The Ga doped $0.01\leq x\leq0.03$ samples undergo insulator to metal
transition at $\sim100K$, while the x = 0.07 sample remains insulating
down to 50K. Both Al and Ga doping exhibits similar magnetic behavior,
though based on their ionic radii they are expected to selectively
substitute $Mn^{4+}$ and $Mn^{3+}$ ions, respectively.

\subsection{Magnetic Structure}

\noindent %
\begin{figure}[t]
\resizebox{0.5\textwidth}{!}{
\includegraphics{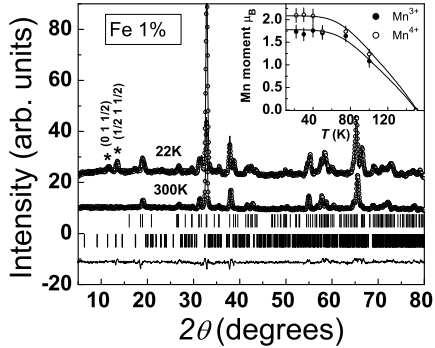}}

\caption{\label{fig:ND_Fe01}Neutron diffraction pattern recorded on sample
$La{}_{0.5}Ca_{0.5}Mn_{0.99}Fe_{0.01}O{}_{3}$ (Fe 1\%) at 300K and
22K is displayed. Data at 22K has been shifted vertically for clarity.
Continuous line through the data points is the fitting to chemical
and magnetic structure described in the text. The ({*}) symbols indicate
superlattice reflections having maximum intensity in the CE-type antiferromagnetic
state. The upper and lower tick marks indicate the indexing of nuclear
and CE-type antiferromagnetic phases, respectively. The curve at the
bottom is difference between observed and calculated intensities at
22K. The inset to figure shows the variation of antiferromagnetic
site moment of $Mn{}^{3+}$ and $Mn{}^{4+}$ ions with temperature
for 1\% Fe doped sample. The continuous lines are a guide for the
eye. }

\end{figure}

\begin{figure}[t]
\resizebox{0.5\textwidth}{!}{
\includegraphics{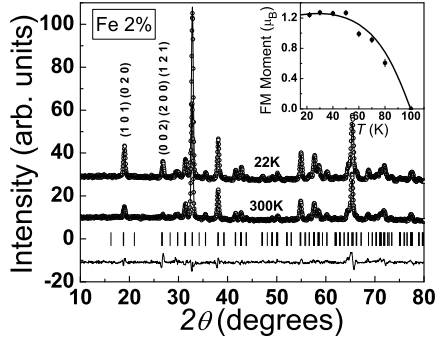}}

\caption{\label{fig:ND_Fe02}Neutron diffraction pattern for $La{}_{0.5}Ca_{0.5}Mn_{0.98}Fe_{0.02}O{}_{3}$
(Fe 2\%) at 22 and 300K. Data at 22K has been shifted vertically for
clarity. Continuous lines are the fitted lines to the chemical and
magnetic structure discussed in the text. The tick marks below the
patterns show the position of the Bragg reflections in \textit{Pnma}
space group. The curve at the bottom is difference between observed
and calculated intensities at 22K. Inset to the figure shows the variation
of ferromagnetic moment with temperature. }

\end{figure}

The neutron diffraction pattern in the angular range $5^{\circ}\leq2\theta\leq80^{\circ}$
of $La{}_{0.5}Ca{}_{0.5}Mn_{0.99}Fe{}_{0.01}O_{3}$ (Fe 1\% doping)
sample at 300 and 22K is shown in figure \ref{fig:ND_Fe01}. This
neutron diffraction pattern is a representative of the samples displaying
CE - type antiferromagnetic spin structure at low temperatures. In
1\% Fe doped sample, on lowering of temperature, below the antiferromagnetic
ordering temperature superlattice reflections are observed similar
to that of $La{}_{0.5}Ca{}_{0.5}MnO_{3}$ compound (figure \ref{fig:ND_LCMO}).
This suggests the antiferromagnetic nature of the sample below the
transition temperature $(T_{N})$. The superlattice reflections are
marked with an asterisks ({*}) symbol in figure \ref{fig:ND_Fe01}.
Particularly, strong reflections (0, 1, \textonehalf{}) and (\textonehalf{},
1, \textonehalf{}) are shown and these characterize the onset of CE
- type antiferromagnetic ordering \cite{P. G. Radaelli}. These superlattice
reflections are indexed on a $2a\times b\times2c$ cell in the space
group $P2{}_{1}/m$. In CE - type antiferromagnetic spin structure
the $Mn{}^{3+}$ and $Mn{}^{4+}$ ions occupy two distinct sites and
are associated with propagation vector (0, 0, \textonehalf{}) and
(\textonehalf{}, 0, \textonehalf{}), respectively. The CE - type model
used for refinement describes the low temperature phase as one - dimensional
zig - zag chains with ferromagnetically aligned spins in the ac plane,
while the interchain coupling is antiferromagnetic. This is the checker
board pattern and the model was first proposed by Wollan and Kohler
\cite{E.O. Wollan}. The temperature dependence of Mn site moment
for $Mn{}^{3+}$ and $Mn{}^{4+}$ ions in 1\% Fe doped compound is
shown in the inset to figure \ref{fig:ND_Fe01}. In 1\% Fe doped compound
$T_{N}\approx150K$ is deduced from the temperature dependence of
the refined Mn site magnetic moment. The Rietveld refinement of the
neutron diffraction pattern at 22K indicates that the magnetic moment
for $Mn{}^{3+}$ and $Mn{}^{4+}$ sites are predominantly oriented
along either a or c axis and their values are 1.9(2) $\mu{}_{B}$
and 2.1(1) $\mu{}_{B}$, respectively. These values are lower in comparison
to the values obtained in the case of x = 0 sample as, 2.8(3)~$\mu{}_{B}$
and 2.6(3)~$\mu{}_{B}$ and that reported by Radaelli et al. \cite{P. G. Radaelli}
for $Mn{}^{3+}$ and $Mn{}^{4+}$ sites, respectively. The lowering
of both the site moment indicates that $Fe^{3+}$ is distributed randomly
over both the sites. $Fe^{3+}$ is a non JT ion like $Cr^{3+}$. However,
unlike $Cr^{3+}$, the number of $e_{g}$ electrons increase with
Fe doping. Our observation are again different from the incommensuration
of $Mn^{3+}$ observed in the case of $Ni^{2+}$ doping \cite{A. Martinelli}.
From the present neutron diffraction data no significant change in
$\chi^{2}$ and magnetic R - factor is observed on changing the orientation
of spins between a and c axis. Therefore, the spins were constrained
to be orientated along a axis. Similarly, the x and z axis component
of magnetic moment of $Mn^{4+}$ ions could not be refined separately
and therefore the orientation was constrained to be along a axis.
Additionally, no enhancement in intensity of low angle fundamental
Bragg reflections was visible, which indicates the absence of ferromagnetic
ordering observable from the present neutron diffraction experiment.
This is further corroborated by SANS data where no signature of ferromagnetism
is observed. In figure \ref{fig:ND_Fe02}, the neutron diffraction
pattern of 2\% Fe doped sample at 22 and 300K is shown. With as little
as 2\% Fe doping, the CE-type antiferromagnetic state is completely
suppressed, indicated by the absence of superlattice reflections at
22K. In addition, significant enhancement in intensity of low angle
nuclear fundamental reflections at 22K is observed. The nuclear reflections
(1 0 1) (0 2 0) and (2 0 0) (0 0 2) (1 2 1) show the maximum enhancement
in intensity. Similar behavior is observed in $La_{0.5}Ca_{0.5}Mn_{1-x}Fe_{x}O_{3}$
samples with $0.02\leq x\leq0.06$. This is a clear indication of
long range ferromagnetic ordering in these samples. The ferromagnetic
phase is fitted in orthorhombic structure in \textit{Pnma} space group.
The ferromagnetic ordering temperatures for 2, 4 and 6\% Fe doped
samples are 100, 150 and 75K, respectively, in concurrence with magnetization
studies. The inset to figure \ref{fig:ND_Fe02} shows the refined
magnetic moment as a function of temperature in 2\% Fe doped sample
with $T{}_{C}\approx100K$. At 22K, the refinement of diffraction
pattern of 2\% Fe doped sample yields the moment value as $\approx1.4(1)\mu{}_{B}$,
in agreement with the M(H) data. The observed magnetic moment is much
lower than the expected value of $\approx3.5\mu{}_{B}$. On further
increasing the Fe doping to 10\% no evidence for the presence of long
range magnetic ordering is observed from neutron diffraction study.
However, the signature for the presence of short range ordered ferromagnetic
phase is evident in M(H) study.

\noindent %

The neutron diffraction pattern of $La{}_{0.5}Ca{}_{0.5}Mn_{1-x}Ru{}_{x}O_{3}$
$(0.01\leq x\leq0.05)$ compounds display characteristics similar
to $La_{0.5}Ca_{0.5}Mn_{0.98}Fe_{0.02}O_{3}$ compound (figure \ref{fig:ND_Fe02}).
The temperature dependent neutron diffraction study on these samples
show a progressive rise in intensity of the low angle fundamental
reflections below $T_{C}\approx200K$. Additionally, no signature
of antiferromagnetic ordering, in form of evolution of superlattice
reflections, are observed. The ferromagnetic phase at low temperature
is refined in orthorhombic structure in \textit{Pnma} space group,
as described for the 2\% Fe doped compound. The refined ferromagnetic
moment at 22K for 1\% and 5\% Ru doped samples are 2.64(5) and 3.19(6)
$\mu{}_{B}/f.u.$, respectively and are in agreement with the M(H) study.

\noindent %
\begin{figure}[t]
\resizebox{0.5\textwidth}{!}{
\includegraphics{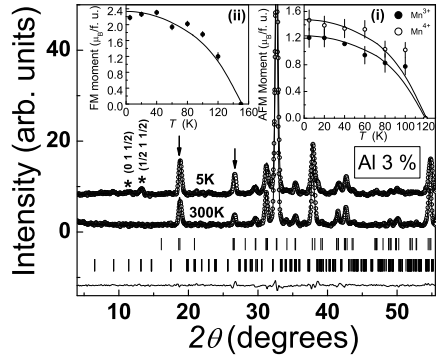}}

\caption{\label{fig:ND_Al03}Neutron diffraction pattern recorded on sample
$La{}_{0.5}Ca_{0.5}Mn_{0.97}Al_{0.03}O{}_{3}$ (Al 3\%) at 300K and
5K is displayed. Data at 5K has been shifted vertically for clarity.
Continuous line through the data points is the fitting to chemical
and magnetic structure described in the text. The ({*}) symbols indicate
superlattice reflections having maximum intensity in the CE-type antiferromagnetic
state and arrow $(\downarrow)$ show the low angle nuclear Bragg reflections
displaying the significant enhancement in intensity at 5K. The upper
and lower tick marks indicate the indexing of nuclear and CE-type
antiferromagnetic phases, respectively. The curve at the bottom is
difference between observed and calculated intensities at 5K. The
inset (i) shows the variation of ferromagnetic moment with temperature
and inset (ii) shows the variation of antiferromagnetic site moment
of $Mn{}^{3+}$ and $Mn{}^{4+}$ ions with temperature for sample
1\% Fe doped sample. The continuous lines are a guide for the eye. }

\end{figure}

\begin{figure}[t]
\resizebox{0.5\textwidth}{!}{
\includegraphics{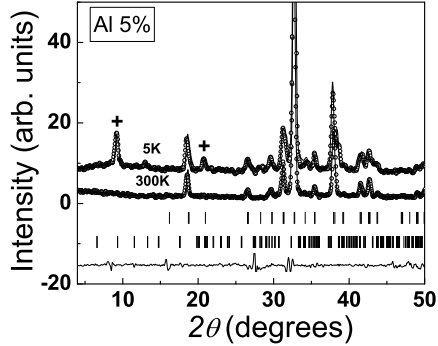}}

\caption{\label{fig:ND_Al05}Neutron-diffraction pattern of $La{}_{0.5}Ca_{0.5}Mn_{0.95}Al_{0.05}O{}_{3}$
(Al 5\%) sample at 5 and 300K. Data at 5K has been shifted vertically
for clarity. Continuous lines through data points are fitted lines
to the chemical and magnetic structure described in the text. The
symbol (+) indicates reflections corresponding to the A - type antiferromagnetic
spin structure. The tick marks in the first and second rows correspond
to positions of Bragg reflections in nuclear and A - type antiferromagnetic
phases, respectively. The curve at the bottom is difference between
observed and calculated intensities at 5K. }

\end{figure}

In the 1\% Al doped sample the superlattice reflections below
$T_{N}\approx150K$, in addition to enhancement in low angle
fundamental Bragg reflections below $T_{C}\approx100K$ are
observed. The superlattice reflections are indexed to CE - type
antiferromagnetic structure. The Rietveld analysis of the
diffraction data is carried out taking both ferromagnetic and
antiferromagnetic phases. The obtained antiferromagnetic moment
values at 5K on $Mn^{3+}$ and $Mn^{4+}$ sites are 2.8(2) and
2.5(2) $\mu{}_{B}$, respectively. The value of antiferromagnetic
moment is close to that observed in the case of x = 0 sample. The
ferromagnetic moment value is 1.1(1) $\mu{}_{B}$ and is in
agreement with the M(H) study. The coexistence of ferromagnetic and
antiferromagnetic phases observed in this sample is distinct from
the Fe doped samples where only CE - type antiferromagnetic state
is evident. In contrast to Fe doped samples, here with non
magnetic Al doping, we find coexisting charge ordered insulating
and ferromagnetic metallic phases. In figure \ref{fig:ND_Al03}, we
have shown the neutron diffraction patterns of
$La_{0.5}Ca_{0.5}Mn_{1-x}Al{}_{x}O_{3}$ with x = 0.03. In this
sample behavior similar to the 1\% Al doped compound is observed.
However, the antiferromagnetic moment values at the respective
$Mn^{3+}$ and $Mn^{4+}$ sites are reduced to 1.2(1) and 1.5(1)
$\mu{}_{B}$, in comparison to the 1\% Al doped compound. The
temperature dependence of the antiferromagnetic moment is shown in the
inset (i) to figure \ref{fig:ND_Al03}. This reduction in CE - type
antiferromagnetic moment is accompanied by the favoring of
ferromagnetic ordering with $T_{C}\approx150K$. The ferromagnetic
moment as a function of temperature in the 3\% Al doped sample is
displayed in the inset (ii) to figure \ref{fig:ND_Al03}. The
refined ferromagnetic moment value at 5K is 2.18(6) $\mu_{B}$,
which is in corroboration with the M(H) study. Therefore, in samples
with $x\leq0.3$, CE-type antiferromagnetic ordering coexists with
the ferromagnetic ordering. This is not very evident from M(T, H)
studies alone. In 5\% Al doped compound, the CE - type
antiferromagnetic state with considerable reduction in moment
value is still evident, whereas the ferromagnetic phase is fully
suppressed. In addition, new distinct superlattice reflections
indicated by (+) in figure \ref{fig:ND_Al05} are observed. The
d-values and the relative intensity of these superlattice
reflections match with those reported previously in half doped
manganites and attributed to A-type antiferromagnetic ordering
\cite{I. Dhiman,H. Kawano,C. Ritter}. The Rietveld refinement of
the diffraction data at 5K shows that these superlattice
reflections correspond to A - type antiferromagnetic ordering.
These superlattice reflections are indexed on $a\times b\times c$
cell in $P\overline{1}$ space group. In the A - type
antiferromagnetic structure Mn spins form ferromagnetic planes
with an antiferromagnetic coupling between them. The magnetic
moment in this A - type antiferromagnetic phase is 2.64(6)
$\mu{}_{B}$, oriented ferromagnetically in ac plane and
antiferromagnetically along b axis. The suppression of CE - type
and emergence of A - type antiferromagnetic state observed in
these systems is similar to that observed in previously reported
neutron diffraction studies on A - site doped compounds \cite{I.
Dhiman,I. Dhiman-1}. On increasing the Al doping to 7\%, both CE -
type antiferromagnetic and ferromagnetic phases are suppressed,
while the superlattice reflections corresponding to an A - type
antiferromagnetic state are still visible below
$T_{N}\approx125K$. The obtained value of magnetic moment at 5K is
3.12(6) $\mu{}_{B}$. Eventually, with increasing Al doping between
10 and 13\%, the A - type antiferromagnetic state is also
suppressed and no signature of long range magnetically ordered
phases are seen from neutron diffraction measurements.

\noindent %

With Ga doping in $La_{0.5}Ca_{0.5}Mn_{1-x}Ga{}_{x}O_{3}$ $(0.01\leq x\leq0.07)$,
the results obtained from the present temperature dependent neutron
diffraction study exhibit behavior similar to Al doped samples. In
low Ga doping of 1\%, the CE-type antiferromagnetic state is observed
below $T_{N}\approx150K$, accompanied by ferromagnetic ordering below
$T_{C}\approx60K$. The refinement of diffraction pattern at 5K yields
antiferromagnetic moment values at $Mn{}^{3+}$ and $Mn{}^{4+}$ sites
as 2.3(2) and 2.3(2) $\mu{}_{B}$, respectively, while the ferromagnetic
moment value is 1.0(1) $\mu{}_{B}$. At 3\% Ga doping also, the CE
- type antiferromagnetic phase is accompanied with the ferromagnetic
state having $T_{C}\approx T_{N}\approx125K$. However in comparison
to the 1\% Ga substituted compound, the CE - type antiferromagnetic phase
is significantly weakened, indicated by the reduced moment values
at $Mn^{3+}$ and $Mn^{4+}$ sites as 1.4(2) and 1.2(1)~$\mu{}_{B}$.
The obtained ferromagnetic moment value at 5K is 2.74(6)$\mu{}_{B}$,
in agreement with the M(H) study. Finally in 7\% Ga doped sample, both
CE - type antiferromagnetic and ferromagnetic states are suppressed,
whereas the A - type antiferromagnetic state persists below $T_{N}\approx75K$.
The obtained moment value at 5K is $\approx1.5(1)\mu_{B}$. The A-type
antiferromagnetic spin structure has been analyzed in a manner similar
to that described for Al doped compounds.

Therefore, the present neutron diffraction measurements together with
magnetization and transport measurements reveal the evolution of various
complex magnetic structures, as a function of temperature and different
B - site dopants. In specific, we find the emergence of A - type antiferromagnetic
spin structure accompanied by the gradual suppression of both CE -
type antiferromagnetic and ferromagnetic phases, on substituting with
non - magnetic dopants in $La_{0.5}Ca_{0.5}Mn_{1-x}B{}_{x}O_{3}$
with B = Al and Ga compounds, which has not been observed before.
This is unlike the behavior reported in $Pr{}_{0.5}Ca{}_{0.5}MnO{}_{3}$
compound, wherein doping with non magnetic ions such as Al, Ga or
Ti leads to the development of pseudo CE - type antiferromagnetic
state at the expense of CE - type antiferromagnetic ground state \cite{C. Martin-1,J. L. Garc=0000EDa-Mu=0000F1oz,C. Yaicle-1}.

\subsection{Small Angle Neutron Scattering}

\noindent %
\begin{figure}[t]
\resizebox{0.5\textwidth}{!}{
\includegraphics{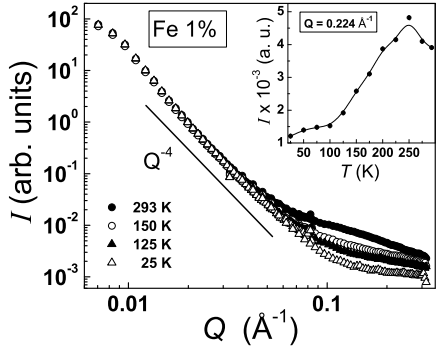}}

\caption{\label{fig:SANS_Fe01}SANS intensity as a function of Q for $La{}_{0.5}Ca_{0.5}Mn_{0.99}Fe{}_{0.01}O{}_{3}$
(Fe 1\%) sample at several temperatures. In the low Q regime the $Q^{-4}$
dependence of intensity is shown. The inset to the figure shows the
SANS intensity as a function of temperature at Q = 0.224$\lyxmathsym{\AA}^{\lyxmathsym{\textminus}1}$
in 1\% Fe doped sample. The continuous line is a guide for the eyes.}

\end{figure}

\begin{figure}[t]
\resizebox{0.5\textwidth}{!}{
\includegraphics{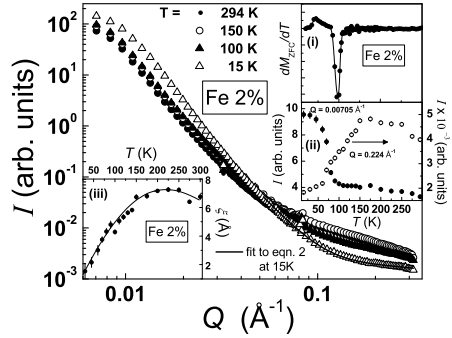}}

\caption{\label{fig:SANS_Fe02}SANS intensity as a function of Q for $La{}_{0.5}Ca_{0.5}Mn_{0.98}Fe{}_{0.02}O{}_{3}$
(Fe 2\%) sample at several temperatures. The continuous line is a
fit to equation \ref{eq:SANS_lorentz} at 15K. The inset (i) displays
the temperature derivative of the ZFC magnetization curve, (ii) shows
the SANS intensity as a function of temperature at Q = 0.00705 and
0.224$\textrm{\AA}^{\lyxmathsym{\textminus}1}$, and (iii) shows the
temperature dependence of correlation length $(\xi)$ in 2\% Fe doped
sample. The continuous line is a guide for the eyes.}

\end{figure}

The presence of short range ordered ferromagnetic correlations as
a result of B - site substitution in $La_{0.5}Ca_{0.5}MnO_{3}$ compound,
for few of the selected samples is explored using SANS measurements,
in the length scale of 10 to 1000$\text{\AA}$. The small angle neutron
scattering (SANS) intensity as a function of Q, I(Q), in the $La_{0.5}Ca_{0.5}Mn_{0.99}Fe_{0.01}O_{3}$
compound (1\% Fe doping) at various temperature is shown in figure
\ref{fig:SANS_Fe01}. Two distinct regimes in $Q$ are observed from
this figure. In the $Q<0.03\lyxmathsym{\AA}^{-1}$ range, the intensity
does not exhibit any temperature dependence, while in the high $Q$
regime significant change in intensity is observed. In the low Q regime,
the intensity exhibits $Q^{-4}$ dependence, obeying classical Porod's
law \cite{G. Porod}. This behavior is attributed to the clustered
structure, where smooth sphere particles are embedded in a matrix
background. On the other hand, in high Q regime i. e. at smaller length
scales, above the ordering temperature $T_{C}\approx230K$, intensity
has contributions from both disordered paramagnetic scattering and
nuclear scattering. On lowering of temperature below the magnetic
ordering, paramagnetic disorder scattering is reduced and only the
nuclear contribution persists as the antiferromagnetic phase does
not contribute to the SANS intensity. The inset to figure \ref{fig:SANS_Fe01}
shows the temperature dependence of scattered neutron intensity at
$Q=0.224\text{\AA}{}^{\lyxmathsym{\textminus}1}$ in $La_{0.5}Ca_{0.5}Mn_{0.99}Fe_{0.01}O_{3}$
compound. For length scale at about $118\text{\AA}$ and beyond, a
maximum at 230K is observed, which coincides with the weak hump in
M(T) (figure \ref{fig:M(T)_R(T)_Fe}(a)). In neutron diffraction measurements
we do not observe any evidence of long range ferromagnetic ordering
in this sample. The decrease in the maximum value of $\rho(T)$ (figure
\ref{fig:M(T)_R(T)_Fe}(b)) in this sample, together with these results
suggests that the reduction in intensity below 230K indicates the
suppression of paramagnetic scattering with the onset of short range
ferromagnetic ordering. However, no further evidence of ferromagnetic
ordering is observed in the low Q regime as observed in other samples.
Therefore, from both neutron diffraction and the SANS study no signature
of ferromagnetic ordering behavior is observed in the zero field CE
- type antiferromagnetic state for 1\% Fe doped sample. Figure \ref{fig:SANS_Fe02}
shows the SANS intensity in $La_{0.5}Ca_{0.5}Mn_{0.98}Fe_{0.02}O_{3}$
(2\% Fe doping) compound at several temperatures between 15 and 295K.
However, in contrast to 1\% Fe doped sample, enhancement in intensity
in the $Q<0.03\lyxmathsym{\AA}^{-1}$ regime is observed on lowering
the temperature below $T_{C}=100K$. The temperature dependence of
SANS intensity at $Q$ = 0.00705 and 0.224$\lyxmathsym{\AA}^{-1}$
for 2\% Fe doped sample is shown in the inset to figure \ref{fig:SANS_Fe02}.
This figure shows the rise in intensity at $Q=0.00705\lyxmathsym{\AA}^{-1}$
below 100K, indicating the ferromagnetic transition, in agreement
with neutron diffraction and M(T) measurements. This behavior correlates
well with the temperature derivative of ZFC magnetization, $dM_{ZFC}/dT$,
wherein the minimum in magnetization close to 100K is evident. While,
in the high Q range at $Q=0.224\lyxmathsym{\AA}^{-1}$, behavior similar
to 1\% Fe doped sample is observed, which is identified as reduction
in paramagnetic disorder scattering accompanied by the onset of short
range ordered ferromagnetic state. The comparison between the SANS
intensity and temperature dependence of strain parameters ($Os_{\shortparallel}$
and $Os_{\perp}$) in 2\% Fe doped sample (inset to figure \ref{fig:Cell_Fe02})
display that the favoring of ferromagnetic ordering is accompanied
by reduction in lattice strain. This behavior is in agreement with
previously reported SANS study on $Pr_{0.7}Ca_{0.3}MnO_{3}$ compound
exhibiting similar magnetic properties, wherein the growth of ferromagnetic
clusters within the antiferromagnetic insulating phase is governed
by reduction in lattice strains in the system \cite{D. Saurel-1}.
In the Q regime between 0.05 to 0.31 $\lyxmathsym{\AA}^{-1}$, the
Q dependence of scattering intensity has been described by the combination
of Porod's law ($Q^{-4}$ dependence) and Lorentzian profile function
($Q^{-2}$ dependence) given as,

\begin{equation}
I=\frac{I_{1}}{Q^{4}}+\frac{I_{2}}{Q^{2}+\kappa^{2}}\label{eq:SANS_lorentz}\end{equation}

\noindent where, $I_{1}$ and $I_{2}$ are scattering amplitudes and
$\kappa=1/\xi$, $\xi$ is the correlation length of magnetic clusters.
The corresponding fits by continuous line at 15K is shown in figure
\ref{fig:SANS_Fe02}. Similar behavior has been reported previously,
wherein the $Q^{-4}$ dependence is ascribed to the scattering by
the phase separation interface, while $Q^{-2}$ dependence arises
from nanometric magnetic inhomogeneities \cite{D. Saurel}. Although,
according to Ch. Simon et al., in $Pr_{0.67}Ca_{0.33}MnO_{3}$ single
crystal sample, the $Q^{-2}$ dependence may be a characteristic of
two dimensional ferromagnetic stripes, suggesting a red cabbage structure
\cite{Ch. Simon}. The correlation length $(\xi)$ as a function of
temperature shown in the inset (iii) to figure \ref{fig:SANS_Fe02}
exhibits a broad maximum between 150 and 250K, which suggests the
presence of short range ordered ferromagnetic clusters even above
the ferromagnetic ordering temperature. In a previous study, we had
observed the evidence of diffuse neutron scattering in the region
$T>T_{C}$, in related $La{}_{0.5}Ca_{0.5-x}Sr{}_{x}MnO{}_{3}$ compounds,
indicating the existence of short range ordered magnetic correlations
in the paramagnetic regime much above the ordering temperatures \cite{I. Dhiman-2}.
The existence of short range ordered ferromagnetic phase is also reported
in many of the SANS measurements on other perovskite manganite systems
\cite{D. Saurel-1,D. Saurel,Ch. Simon} and is a signature of magnetic
polarons. The scattering amplitude $I_{1}$ follows the temperature
dependence of intensity at $Q=0.00705\lyxmathsym{\AA}^{-1}$ shown
in the inset to figure \ref{fig:SANS_Fe02}, while no significant
change in $I_{2}$ as a function of temperature is observed.

\noindent %

\begin{figure}[t]
\resizebox{0.5\textwidth}{!}{
\includegraphics{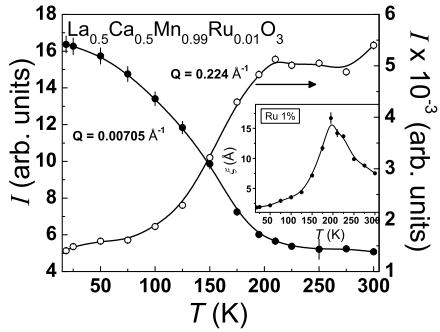}}

\caption{\label{fig:IQ_Ru01}SANS intensity as a function of temperature at
Q = 0.00705 and 0.224$\lyxmathsym{\AA}^{\lyxmathsym{\textminus}1}$
in $La{}_{0.5}Ca_{0.5}Mn_{0.99}Ru{}_{0.01}O{}_{3}$ sample. The inset
exhibits the temperature dependence of correlation length $(\xi)$
in 1\% Ru doped sample. The continuous line is a guide for the eyes.}

\end{figure}

In figure \ref{fig:IQ_Ru01}, the temperature dependence of the SANS
intensity at $Q$ = 0.00705 and 0.224$\lyxmathsym{\AA}^{-1}$ for
$La_{0.5}Ca_{0.5}Mn_{0.99}Ru{}_{0.01}O_{3}$ (1\% Ru doping) sample
is shown. In this sample, the temperature dependence of SANS intensity
as a function of Q, exhibits behavior similar to 2\% Fe doped compound
(figure \ref{fig:SANS_Fe02}). In the low Q regime $(0.00705\leq Q\leq0.033\lyxmathsym{\AA}^{-1})$,
the scattering intensity follows a $Q^{-4}$ dependence, whereas at
higher Q values the behavior is described by equation \ref{eq:SANS_lorentz}.
The obtained correlation length as a function of temperature, shown
in the inset to figure \ref{fig:IQ_Ru01}, exhibits a sharp maximum
across the ferromagnetic ordering temperature. The temperature of
such a feature correlates well with one obtained from the magnetization
measurement $(T_{C}\approx200K)$. At $Q=0.00705\lyxmathsym{\AA}^{-1}$,
the intensity as a function of temperature, shown in figure \ref{fig:IQ_Ru01},
displays an increase on cooling below 200K, indicating the ferromagnetic
nature of the sample. This behavior is followed by an increase in
SANS intensity on raising the temperature above 200K, at smaller $Q=0.224\lyxmathsym{\AA}^{-1}$,
evident in the figure \ref{fig:IQ_Ru01}. This behavior also correlates
with the minimum observed in $dM_{ZFC}/dT$ plot. In this 1\% Ru doped
compound also, similar to 2\% Fe doping, evidence of ferromagnetic
ordering on two different length scales is observed.

\noindent %

\begin{figure}[t]
\resizebox{0.5\textwidth}{!}{
\includegraphics{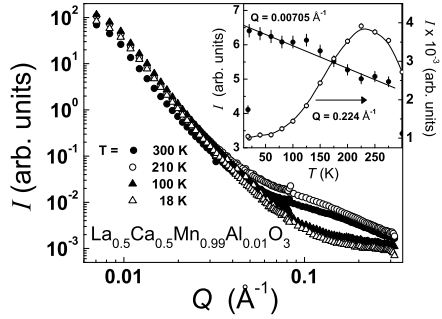}}

\caption{\label{fig:SANS_Al01}SANS intensity as a function of Q for $La{}_{0.5}Ca_{0.5}Mn_{0.99}Al{}_{0.01}O{}_{3}$
(Al 1\%) sample at several temperatures. The inset shows the SANS
intensity as a function of temperature at Q = 0.00705 and 0.224$\lyxmathsym{\AA}^{\lyxmathsym{\textminus}1}$
in 1\% Al doped sample. The continuous line is a guide for the eyes.}

\end{figure}

The variation of small angle neutron scattering intensity as a function
Q, at several temperatures between 18 and 300K, for $La_{0.5}Ca_{0.5}Mn_{0.99}Al{}_{0.01}O_{3}$
(1\% Al doping) sample is shown in figure \ref{fig:SANS_Al01}. The
neutron diffraction measurements on 1\% Al doped sample shows two
transitions at $T_{N}\approx150K$ and $T_{C}\approx100K$. In figure
\ref{fig:SANS_Al01}, for low Q values at low temperatures, the crossing
over of SANS curves is found. Although, enhancement in intensity at
low Q values below 0.044$\lyxmathsym{\AA}^{-1}$ is considerably small,
as compared to 2\% Fe and 1\% Ru compounds. The temperature variation
of intensity at $Q=0.00705\lyxmathsym{\AA}^{-1}$ exhibits almost
linear increase upon cooling below 300K (the inset to figure \ref{fig:SANS_Al01}).
At high Q value of 0.224$\lyxmathsym{\AA}^{-1}$, the scattering intensity
displays a sharp peak at 200K, coinciding with the broad hump in M(T)
data. This indicates the existence of nanometric size ferromagnetic
regions, which coexists with the long ranged antiferromagnetic phase
below $T_{N}\approx150K$. SANS measurements have also been performed
on higher Al doped sample in $La_{0.5}Ca_{0.5}Mn_{0.95}Al{}_{0.05}O_{3}$
(5\% Al doping). Contrastively, this sample exhibits the stabilization
of A - type antiferromagnetic state at low temperatures. The temperature
variation of SANS intensity I(Q) for 5\% Al doped sample displays
characteristics similar to 1\% Fe doped compound (figure \ref{fig:SANS_Fe01}).
At low Q (< 0.055$\lyxmathsym{\AA}^{-1}$) values the scattering curves
overlap, displaying no significant change with temperature. While,
in the high Q regime at $Q=0.224\lyxmathsym{\AA}^{-1}$ scattering
intensity exhibits a decrease below 200K, attributed to the reduction
of paramagnetic disorder scattering accompanied with the growth of
nanometric sized ferromagnetic regions.

\noindent %

The above results clearly demonstrate the contrastive nature of magnetic
(Fe and Ru) and non magnetic dopants (Al and Ga) in half doped $La_{0.5}Ca_{0.5}MnO_{3}$
compound. According to previous studies on B - site substituted charge
ordered manganites, contrasting nature in the stabilization of magnetic
ground states has been ascribed to different mechanisms, such as the
valence state of dopants, magnetic exchange interaction with Mn ions,
and between dopants plays an essential role. These studies have shown
that Fe ions couple antiferromagnetically with Mn ions \cite{S. B. Ogale,L. K. Leung,A. Simopoulos},
while Ru ions couple ferromagnetically with Mn ones \cite{C. Martin,V. Markovich-1,V. Markovich-2,J. S. Kim}.
In addition, Ru doping enhances the local $e_{g}$ electron density;
therefore strongly favor ferromagnetic metallic phase \cite{C. L. Lu}.
These effects may together contribute towards the favoring of insulator
to metal transition in Ru doped samples \cite{X. Chen-1,K. Pradhan}.
On the other hand, the non magnetic dopants are considered as random
impurities, not having any magnetic coupling with Mn ions. The effect
of random substitution at the B - site by magnetic and non-magnetic
dopants has been investigated by Monte Carlo simulation studies in
half doped manganites \cite{X. Chen-1,K. Pradhan,C. L. Lu}. The non
- magnetic impurity doping at the B - site is introduced as lattice
defects and may modify the $e_{g}$ electron density, while in the
case of magnetic doping, the magnetic exchange interactions are also
taken into account. Our experimental studies are fairly consistent
with these theoretical studies, wherein the long range ordered antiferromagnetic
state collapses into a ferromagnetic phase and this behavior is ascribed
to lattice defects introduced in the form of non magnetic B - site
dopants. However, our studies indicate the absence of proposed C -
type antiferromagnetic state \cite{X. Chen-1}, and instead the presence
of A - type antiferromagnetic state in the vicinity of melting of
CE - type antiferromagnetic and ferromagnetic phase is observed. This
behavior has been predicted theoretically by Pradhan et al. for B
site dopants with 3+ valence, where phase separation window lies between
charge ordered CE - type and A - type antiferromagnetic phase \cite{K. Pradhan}.
Summarizing, our experimental studies show that the charge and orbitally
ordered antiferromagnetic state can be significantly destabilized
by the induced B - site disorder, leading to variation in favoring
of magnetic ground states with different dopants. These include competing
long range ordered ferromagnetic metallic or insulating state, short
range ordered ferromagnetic phase coexisting with or without the antiferromagnetic
state, or a system may enter magnetically frustrated, spin glass like
state. In particular, the development of an A - type antiferromagnetic
tendencies observed here are driven by the non magnetic substitution
(Al or Ga) in the $La{}_{0.5}Ca{}_{0.5}MnO_{3}$ compound. The growth
of ferromagnetic phase for low concentration of B - site dopants and
its suppression at higher concentration is understood to be a combined
effect of defects induced and density driven phenomena.

\section{Conclusions}

The influence of B - site doping on the crystal and magnetic structure
in $La_{0.5}Ca_{0.5}Mn_{1-x}B_{x}O_{3}$ (B = Fe, Ru, Al and Ga) has
been investigated using neutron diffraction, SANS, magnetization and
resistivity techniques. The B - site doped samples are isostructural
and crystallize in an orthorhombic structure in \textit{Pnma} space
group at 300K. On lowering of temperature, orthorhombic to monoclinic
structural transition is observed in compounds exhibiting CE - type
antiferromagnetic ordering. This structural transition is absent in
other compounds which exhibit A - type antiferromagnetic and ferromagnetic
ordering. In 1\% Fe doped compound CE - type antiferromagnetic and
ferromagnetic ordering is found similar to the x = 0 sample. However,
the moment on both $Mn^{3+}$ and $Mn^{4+}$ are reduced. With higher
Fe doping, CE - type antiferromagnetic state is suppressed at the
expense of ferromagnetic insulating phase in $0.02\leq x\leq0.06$
compounds. At higher Fe doping in $x>0.06$, the ferromagnetic state
is also suppressed, no evidence of long range magnetic ordering is
found. Contrastively, Ru doping as low as x = 0.01 favors ferromagnetic
metallic state at $T_{C}\approx200K$ and $T_{MI}\approx125K$. With
Al substitution in $0.01\leq x\leq0.03$, the charge ordered CE -
type antiferromagnetic state is retained, coexisting with ferromagnetic
metallic phase. On increasing the Al doping $(x>0.03)$, both CE -
type antiferromagnetic and ferromagnetic phases are gradually suppressed
and an A - type antiferromagnetic insulating state is stabilized in
$0.05\leq x\leq0.07$, which is also suppressed eventually at higher
doping. On doping with Al, the charge and orbitally ordered state
is retained up to much higher doping level with x = 0.07. Similarly,
substitution with Ga is observed to induce similar effects as described
for Al doped samples. The contrasting behavior of magnetic Fe and
Ru as against non magnetic dopants Al and Ga are evidenced from this
study. Doping with Fe and Ru is found result in either ferromagnetic
or antiferromagnetic phases. While, doping with Al and Ga results
in coexisting ferromagnetic and antiferromagnetic phases.

\section{Acknowledgement}

One of the authors (AD) gratefully acknowledges Department of
Science and Technology (DST), India for financial support to carry
out the SANS experiments at PSI, Switzerland.

\end{document}